\documentclass{aa}  

\usepackage{graphicx}
\usepackage{txfonts}

\usepackage{color}

\begin{document} 

\title{Does the mean-field alpha effect have any impact on the memory of the solar cycle?}

\author{Soumitra Hazra$^{1,2}$\and Allan Sacha Brun$^1$ \and Dibyendu Nandy$^{3,4}$}

\institute{
Département d'Astrophysique/AIM, CEA/IRFU, CNRS/INSU, Université Paris-Saclay, Université de Paris, CEA Paris-Saclay, Bat 709, 91191 Gif-sur-Yvette, France\\
Institut d’Astrophysique Spatiale, CNRS, Univ. Paris-Sud, Université Paris-Saclay, Bât. 121, F-91405 Orsay, France\and
Department of Physical Sciences, Indian Institute of Science Education and Research Kolkata,  Mohanpur 741246, West Bengal, India\and
Center of Excellence and Space Sciences India, Indian Institute of Science Education and Research Kolkata, Mohanpur 741246, \\
West Bengal, India\\
}   


\titlerunning{Impact of mean-field alpha effect on solar cycle memory}

\authorrunning{Hazra, Brun \& Nandy}

  \abstract
   {Predictions of solar cycle 24 obtained from advection-dominated and diffusion-dominated kinematic dynamo models are different if the Babcock-Leighton mechanism is the only source of the poloidal field. Yeates et al. (2008) argue that the discrepancy arises due to different memories of the solar dynamo for advection- and diffusion-dominated solar convection zones.}
   {We aim to investigate the differences in solar cycle memory obtained from advection-dominated and diffusion-dominated kinematic solar dynamo models. Specifically, we explore whether inclusion of Parker's mean-field $\alpha$ effect, in addition to the Babcock-Leighton mechanism, has any impact on the memory of the solar cycle.}
   {We used a kinematic flux transport solar dynamo model where poloidal field generation takes place due to both the Babcock-Leighton mechanism and the mean-field $\alpha$ effect. We additionally considered stochastic fluctuations in this model and explored cycle-to-cycle correlations between the polar field at minima and toroidal field at cycle maxima.}
   {Solar dynamo memory is always limited to only one cycle in diffusion-dominated dynamo regimes while in advection-dominated regimes the memory is distributed over a few solar cycles. However, the addition of a mean-field $\alpha$ effect reduces the memory of the solar dynamo to within one cycle in the advection-dominated dynamo regime when there are no fluctuations in the mean-field $\alpha$ effect. When fluctuations are introduced in the mean-field poloidal source a more complex scenario is evident, with very weak but significant correlations emerging across a few cycles.}
   {Our results imply that inclusion of a mean-field $\alpha$ effect in the framework of a flux transport Babcock-Leighton dynamo model leads to additional complexities that may impact memory and predictability of predictive dynamo models of the solar cycle.}


\maketitle
   
\section{Introduction}
The magnetic field of the Sun is responsible for most of the dynamical features in the solar atmosphere. The solar cycle is the most prominent signature of solar magnetic activity in which the number of sunspots, which are strongly magnetized regions on the solar surface, varies cyclically with a periodicity of 11 years. When the Sun reaches the peak of its activity cycle, there are a large number of flares and coronal mass ejections, which can affect vulnerable infrastructures of our modern society \citep{schr15}. These important issues highlight the need for solar activity predictions, which will enable us to mitigate the impact of our star's active behaviour \citep{hath09, petr10}. In recent years, many theoretical and observational studies have been performed to predict solar activity, but the results are diverging \citep{pesn08}.

Our current understanding of the solar cycle suggests that sunspots originate from the buoyant emergence of toroidal flux tubes which are generated via the dynamo mechanism inside the solar interior. The dynamo mechanism involves the joint generation and recycling of the toroidal and the poloidal components of the solar magnetic field \citep{park55}. Pre-existing poloidal magnetic field components are stretched along the $\phi$-direction due to strong differential rotation, generating the toroidal magnetic field. It is thought that toroidal field generation takes place throughout the solar convection zone, but is amplified near the base of the convection zone. Tachocline, a region of strong radial gradient in rotation and low diffusivity, offers an ideal location for storage and amplification of the toroidal magnetic field. Sufficiently strong toroidal flux tubes become magnetically buoyant and emerge at the solar surface in the form of sunspots. However, two different proposals exist in the literature for the poloidal field generation--one involves the decay and dispersal of bipolar magnetic regions at the solar surface, termed as the Babcock-Leighton mechanism \citep{babc61, leig69} and the other evokes strong helical turbulence inside the solar convection zone, known as the mean-field alpha effect \citep{park55, stee66}.  In recent years, dynamo models based on the Babcock-Leighton mechanism have been successful in explaining different observational aspects regarding solar activity \citep{dikp99, nand02, chou04, jouv07, nand11, chou12, bhow18, shaz19, bhow19}. Recently, data-driven 2.5D kinematic dynamo models and 3D kinematic solar dynamo models have also been developed to study different observational aspects regarding solar activity \citep{brun07, jouv11, yeat13, hung17, ghaz17, kara17, ghaz18, kuma19}. For reviews of the solar and stellar dynamo model, see \cite{char05}, \cite{brun15}, and \cite{brun17}. 

As there is a spatial separation between the source layers of the toroidal and poloidal field,  there must be some effective communication mechanism between these layers. While magnetic buoyancy plays a primary role in transporting the toroidal flux from the base of the convection zone to the solar surface, alternative flux transport mechanisms, namely diffusion, meridional flow, and turbulent pumping, share the role of transporting the poloidal flux from the surface to the base of the convection zone. It has been shown that there is a finite time required for the magnetic flux transport which impacts the predictability of the solar cycle \citep{yeat08, jouv10}. \cite{dikp06} used an advection-dominated dynamo model (where the meridional flow is the primary flux transport mechanism) to predict solar cycle 24 and found that cycle 24 should have been a strong one. We note that \cite{dikp06} used a weak tachocline alpha effect in their model. However, \cite{chou07} used a diffusion-dominated dynamo model (diffusion is the primary flux transport mechanism) to predict solar cycle 24, which led to a prediction that cycle 24 will be a weaker one. \cite{yeat08} showed that the memory of the solar cycle in the diffusion-dominated dynamo is shorter (only one cycle) while the memory of the solar cycle in advection-dominated dynamo lasts over a few solar cycles. These latter authors suggest that the difference in the memory of the solar cycle in the two regimes results in different predictions of the solar cycle. Multi-Cycle memory in the advection-dominated dynamo indicates that poloidal fields of the cycle n-1, n-2 and n-3 combine to generate the toroidal field of cycle n. On the other hand, one cycle memory in the diffusion-dominated dynamo suggests that only poloidal field of cycle n-1 is responsible for the generation of the toroidal field of cycle n. Later, \cite{kara12} showed that the introduction of turbulent pumping reduces the memory of the solar cycle to one cycle in both advection and diffusion-dominated dynamo models, which impacts the capability of these kinds of models for prediction.  Turbulent pumping transports the magnetic field vertically downwards; however, there is also a significant latitudinal component in the strong rotation regime \citep{osse02, kap06b, kap06a, mass08, doca11, shaz16}.
   
Most of the dynamo-based prediction models completely ignore the contribution of distributed mean-field alpha effect; they consider the Babcock-Leighton mechanism as the only poloidal-field-generation mechanism for their prediction models. However, some studies indicate that the mean-field alpha effect plays an important role in solar dynamo models and is necessary to recover the solar cycle from grand-minima-like episodes \citep{pipi11, pipi13, pass14, shaz14, ince19}. Recently, \cite{bhow18} considered both the Babcock-Leighton mechanism and mean-field alpha effect as poloidal-field-generation mechanisms in their model to predict the strength of solar cycle 25. Here, we want to explore the importance of the mean-field alpha effect in the context of solar cycle memory and predictability.  We find that the presence of mean-field alpha reduces the memory to one cycle for both advection- and diffusion-dominated regimes. We provide details about our solar dynamo model in Section 2 followed by a discussion of our results in Section 3. Finally, in the last section, we present our conclusions.

\section{Model}
Our ($\alpha-\Omega$) kinematic solar dynamo model solves the evolution equations for the toroidal and poloidal components of solar magnetic fields \citep{moff78, char05}:
\begin{equation}\label{Eq_2.5DynA}
    \frac{\partial A}{\partial t} + \frac{1}{s}\left[ \textbf{v}_p \cdot \nabla (sA) \right] = \eta\left( \nabla^2 - \frac{1}{s^2}  \right)A + S_{p}
,\end{equation}\\
 
\begin{eqnarray}\label{Eq_2.5DynB}
    \frac{\partial B}{\partial t}  + s\left[ \textbf{v}_p \cdot \nabla\left(\frac{B}{s} \right) \right] 
    + (\nabla \cdot \textbf{v}_p)B = \eta\left( \nabla^2 - \frac{1}{s^2}  \right)B  \nonumber \\ 
    + s\left(\left[ \nabla \times (A\bf \hat{e}_\phi) \right]\cdot \nabla \Omega\right)   
    + \frac{1}{s}\frac{\partial (sB)}{\partial r}\frac{\partial \eta}{\partial r}~~~~~
,\end{eqnarray}

where, $s = r\sin(\theta)$ and $v_p$ is the meridional flow. We specify the differential rotation and turbulent magnetic diffusivity by $\Omega$ and $\eta,$ respectively. Here, $B$ represents the toroidal magnetic field components and $A$ represents the vector potential of the poloidal magnetic field component.   In the poloidal field evolution equation, $S_{p}$ is the source term for the poloidal field; while the second term in the RHS of the toroidal field evolution is the source term for the toroidal field due to differential rotation.

We do not consider small-scale convective flows in this model. However, we consider an effective turbulent diffusivity in our model to capture the mixing effects due to convective flows. We do not have a reliable estimate of the diffusivity value inside the convection zone at this moment. However, the diffusivity value near the surface is well constrained by surface flux transport dynamo models; as well as by observations \citep{komm95, muno11, leme15}. Diffusivity values near the surface have been found to be a few times $10^{12}$ cm$^2$/s. It is still unclear how these surface values change as a function of depth in the solar convection zone. We assume a profile that keeps a value that  close to value at the surface, except in the tachocline where it drops by several orders of magnitude due to the reduced level of turbulence there. Recent theoretical studies also suggest a diffusivity value of the order of $10^{12}$ cm$^2$/s inside the convection zone \citep{park79, mies12, came16}. 
 We use a two-step radial diffusivity profile that has the following form: 
\begin{eqnarray}
 \eta(r) = \eta_{bcd} + \frac{\eta_{cz} - \eta_{bcd}}{2}\left( 1 + \operatorname{erf}\left( \frac{r - r_{cz}}{d_{cz}}  \right)
      \right) \nonumber \\
      ~~~~~~~~~~~+ \frac{\eta_{sg} - \eta_{cz} - \eta_{bcd}}{2}\left( 1 + \operatorname{erf}\left( \frac{r - r_{sg}}{d_{sg}}  \right) \right),
\end{eqnarray}
 where $\eta_{bcd}= 10^8$ cm$^2$/s is the diffusivity at the bottom of the computational domain, $\eta_{cz} = 10^{12}$ cm$^2$/s is the diffusivity in the convection zone, and $\eta_{sg} = 2 \times 10^{12}$ cm$^2$/s is the near surface supergranular diffusivity. Other parameters, which characterize the transition from one value of diffusivity to another, are taken as $r_{cz} = 0.73R_\odot$, $d_{cz} = 0.015R_\odot$, $r_{sg} = 0.95R_\odot$, and $d_{sg} = 0.015R_\odot$. 
 
We use an analytic fit to the observed helioseismic rotation data as our differential rotation profile (see \cite{nand11, munoz09}):
\begin{equation}\label{DRan}
   \begin{array}{cc}
      \Omega(r,\theta) = 2\pi\Omega_{c} + \pi \left( 1 - \operatorname{erf}\left( \frac{r - r_{tc}}{d_{tc}}  \right)\right) 
      \left( \Omega_{e} - \Omega_{c} + ( \Omega_{p} - \Omega_{e} )\Omega_S(\theta) \right) ,\\
      \\
      \Omega_S(\theta) = a\cos^2(\theta) + (1-a)\cos^4(\theta), \\
    \end{array}
\end{equation}
 where $\Omega_{c}$, $\Omega_{e}$, and $\Omega_{p}$ represent the rotation frequencies of the core, equator, and the pole, respectively. We take $\Omega_{c} = 432$ nHz, $\Omega_{e} = 470$ nHz, $\Omega_{p} = 330$ nHz,  $r_{tc} = 0.7R_\odot$, $d_{tc} = 0.025R_\odot$ (half of the tachocline thickness), and $a = 0.483$.

Recent helioseismic results have not yet converged to provide an accurate picture of the structure of the meridional flow \citep{raja15, jack15, zhao16}. As information about the meridional flow structure is absent at present moment, we use a single-cell meridional circulation (${\bf v}_p$) profile which transports the field poleward at the surface and equatorward at the base of the convection zone; see \cite{jouv07}, \cite{ghaz14}, and \cite{shaz16} for a discussion on the role of multicellular or shallow meridional flow profiles. We obtained the profile for the meridional circulation ($v_p$) for a compressible flow inside the convection zone using the following equation:
\begin{equation}
  \nabla.(\rho v_p)=0.
 \end{equation}
  So,
  \begin{equation}
  \rho v_p = \nabla \times (\psi \hat{e_\phi}),
  \end{equation}
where $\psi$ is prescribed as:
  \begin{eqnarray}
   \psi r \sin \theta =   
   \psi_0 (r - R_p) \sin \left[ \frac{\pi (r - R_p)}{(R_\odot - R_p)} \right] \{ 1 - e^{- \beta_1 r \theta^{\epsilon}}\} \nonumber \\
  \times~~  \{1 - e^{\beta_2 r (\theta - \pi/2)} \} e^{-((r -r_0)/\Gamma)^2}
,\end{eqnarray}
 where $\psi_0$ controls the maximum speed of the flow. We take the following parameter values to obtain the profile for meridional circulation: $ \beta_1=1.5, \beta_2=1.8, \epsilon=2.0000001, r_0=(R_\odot-R_b)/4, \Gamma=3.47 \times 10^8 , \gamma=0.95, m=3/2$. Here, $R_p=0.65R_\odot$ corresponds to the penetration depth of the meridional flow, and $R_b = 0.55 R_\odot$ is the bottom boundary of our computational domain. Both observation of small-scale features on the solar surface and helioseismic inversions indicate that the surface flow from the equator to the pole has an average speed of 10-25 ms$^{-1}$ \citep{komm93, snod96, hath96}. In our model, the meridional flow speed at the surface lies within the range of 10-25 ms$^{-1}$ and reduces to 1 ms$^{-1}$ at the base of the convection zone.
 
To explore the importance of the mean-field alpha effect, we consider two distinct scenarios. In the first, poloidal field generation takes place only due to the Babcock-Leighton mechanism; while in the second scenario poloidal field alpha generation takes place due to the combined effect of the Babcock-Leighton mechanism and mean-field alpha effect.  In the first scenario, $S_p = S_{BL}$, where $S_{BL}$ is the source term for the poloidal field due to the Babcock-Leighton mechanism. We model the poloidal field source term due to the Babcock-Leighton mechanism by the methods of a double ring first proposed by \cite{durn97}. Subsequently, other groups used the double-ring algorithm to model the Babcock-Leighton mechanism in their dynamo models \citep{nand01, muno10, nand11, shaz13, shaz16}. It has been shown that the double-ring algorithm captures the essence of the Babcock-Leighton mechanism in a better way compared to other formalisms \citep{muno10}. We provide details of our double-ring algorithm in the Appendix. Please note that when we model the Babcock-Leighton mechanism via the double-ring algorithm, we fix the Babcock-Leighton source term in the equation (1) to zero, and we modify the poloidal field by the poloidal fields associated with the double ring (i.e. $A(i,j)$ is modified by $A(i,j) + A_{doublering}$) at regular time intervals. As the double-ring algorithm works above a certain threshold, a recovery mechanism is necessary to recover an activity level of the Sun from grand-minima-like phases. However, some previous studies indicate that even if there are no sunspots during grand minima, there are still many ephemeral regions at the solar surface which obey the Hale's polarity law. These ephemeral regions may contribute to the poloidal-field-generation mechanism during this time \citep{prie14, svan16, kara18}. Therefore, we also added an extra Babcock-Leighton source term due to ephemeral regions which acts on the weak magnetic field regime; see the Appendix for details of the Babcock-Leighton source terms. In this way, we ensure the effectiveness of the Babcock-Leighton mechanism throughout our simulation.
  
 In the second Scenario, $S_p = S_{BL} + S_{MF}$, where $S_{MF}$ is the poloidal field source term due to mean-field alpha effect. This implies that the poloidal field is generated due to both the Babcock-Leighton mechanism and mean-field alpha effect. We model the mean-field alpha effect following this equation:
 \begin{eqnarray}
    S_{MF}= S_{\odot} \frac{ \cos \theta }{4} \left[1+\textrm{erf}
    \left( \frac{r-r_1}{d_1}\right)\right]  
    \left[1-\textrm{erf}\left( \frac{r-r_2}{d_2}\right)\right] \nonumber \\ 
    \times \frac{1}{1+\left(\frac{B_\phi}{B_{up}}\right)^2} ~~~~
   ,\end{eqnarray}
 where $r_1=0.71 R_0$, $r_2=R_0$, $d_1=d_2=0.25 R_0$, and $B_{up}= 10^4~G,$ which is the upper threshold. Here, $S_\odot$ controls the amplitude of the mean-field alpha effect.The function $1/(1+\left(\frac{B_\phi}{B_{up}}\right)^2)$ ensures that this additional $\alpha$ effect is only effective on weak magnetic field strength (below the upper threshold $B_{up}$) and the values of $r_1$ and $r_2$ ensure that this additional mechanism takes place inside the bulk of the convection zone (see top panel of Fig.~1 for radial profile of the mean-field $\alpha$-coefficient). We set the critical value of $S_\odot$ such that our model generates periodic cycles if we consider the mean-field alpha effect as the only poloidal-field-generation mechanism. The critical value of $S_\odot$ is 0.14 m s$^{-1}$ for our model. Please see the right-hand side of the upper panel for the radial profile of the mean-field alpha coefficient.
 
 We perform all of our dynamo simulations within the meridional slab $0.55 R_\odot < r < R_\odot$ and $0 < \theta < \pi$ with a resolution of $300 \times 300$ (i.e. $N_r =N_\theta =300$). We set $A=0$ and $B \propto \sin(2 \theta) \sin(\pi ((r-0.55 R_\odot)/(R_\odot-0.55 R_\odot)))$ as dipolar initial conditions for our simulations. Finally, we solve the dynamo equations with proper boundary conditions suitable for the Sun. As our model is axisymmetric, we set both poloidal and toridal fields at zero 
 ($A=0$ and $B=0$) at the pole ($\theta=0$ and $\theta=\pi$) to avoid any kind of singularity. The inner boundary condition at the bottom of the computational domain ($r=0.55 R_\odot$) is of a perfect conductor. Therefore, at $r=0.55 R_\odot$, both the toroidal and poloidal field components vanish (i.e. $ A=0$ and $B=0$). We assume that there is only the radial component of the solar magnetic field at the surface, which is necessary for stress balance between the subsurface and coronal magnetic fields \citep{vanb07}. We set $B=0$ and $\partial (rA)/\partial r =0$ as a top boundary condition at the surface ($r=R_\odot$).
\begin{figure*}
  \begin{center}
\begin{tabular}{cc}
\includegraphics[scale=0.5]{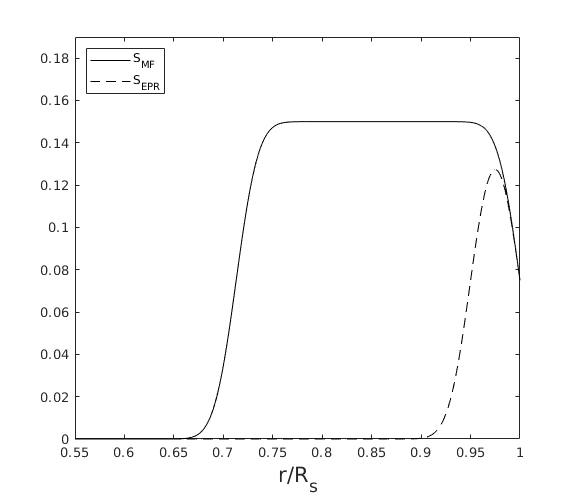} 
\includegraphics[scale=0.5]{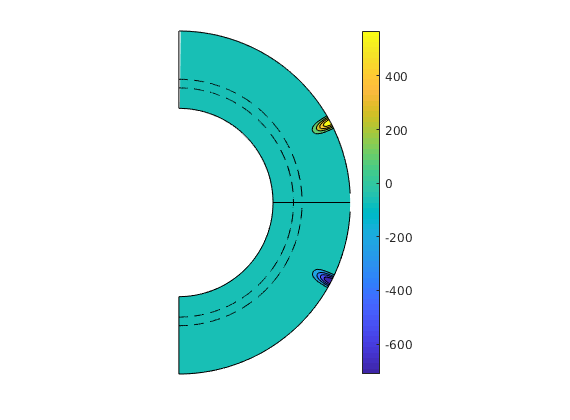} \\
\includegraphics[scale=0.9]{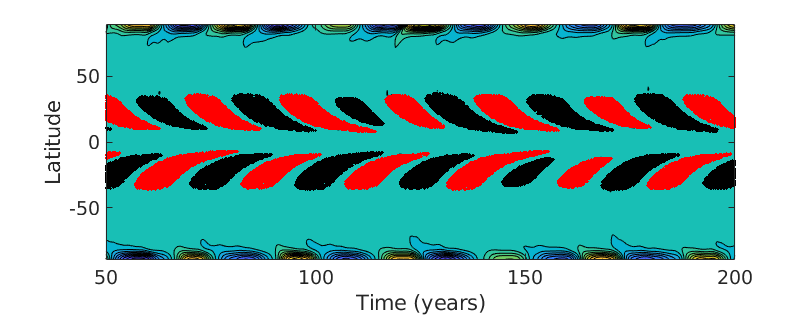}\\
\includegraphics[scale=0.7]{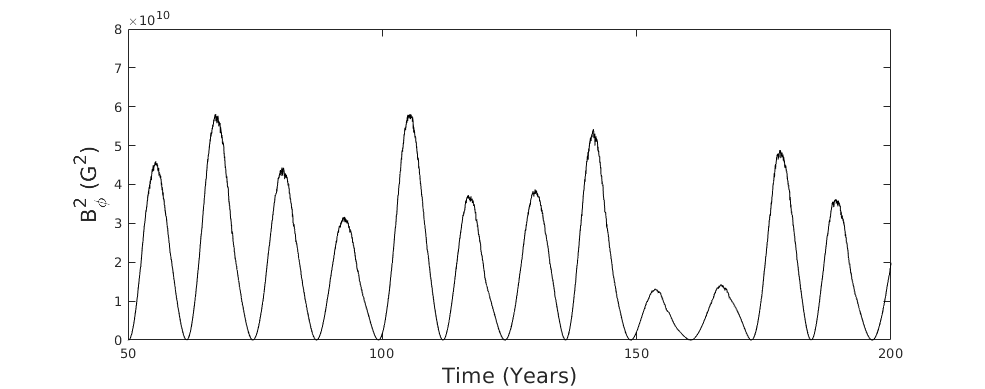}
\end{tabular}
\end{center}  
\caption{ \footnotesize Top panel: (left) Radial profile of mean-field $\alpha$-coefficient and the Babcock-Leighton source term due to ephemeral regions. (right) Poloidal field line contours obtained from the double-ring algorithm in the northern and southern hemispheres, respectively. Middle panel: Butterfly diagram generated from our simulation in the diffusion-dominated region. Bottom panel: Typical variation of $B_\phi^2$ at the base of the solar convection zone with time.}
\end{figure*}

\section{Results}
In order to constrain the impact of the mean-field alpha effect on the memory of the solar cycle, we perform kinematic solar dynamo simulations in two different regimes: advection dominated ($\eta_{cz} = 1 \times 10^{12}$ cm$^2$ s$^{-1}$, $v_0 = 25$ m s$^{-1}$) and diffusion dominated   ($\eta_{cz} = 1 \times 10^{12}$ cm$^2$ s$^{-1}$, $v_0 = 15$ m s$^{-1}$). Advection-dominated regimes are characterised by the dominance of meridional circulation as a major poloidal flux transport mechanism from the surface to the base of the convection zone, while diffusion-dominated regimes are characterised by the dominance of diffusion \citep{yeat08, kara12}. We define the advective flux transport timescale following the suggestions of \cite{yeat08}. The advective flux transport timescale is the time taken for the meridional circulation to transport poloidal fields from $r=0.95~R_\odot$, $\theta=45 ^{\circ}$ to the location at the tachocline where the strongest toroidal field is formed ($\theta=60^{\circ}$). The meridional flow speed of the order of $25$ m s$^{-1}$ at the surface yields an advection flux transport timescale of about 9-10 years (with a flow speed of $15$ m s$^{-1}$ this becomes 16 years). While turbulent diffusion of the order of $1 \times 10^{12}$ cm$^2$ s$^{-1}$ gives us the diffusion flux transport timescale ($L^2/\eta$ where $L$ is the depth of the convection zone) of 14 years. In the diffusion-dominated regime, the diffusion timescale is shorter than the advection timescale, and vice versa in the advection-dominated regime. 

\begin{figure*}
        \centering
        \includegraphics[width=15 cm]{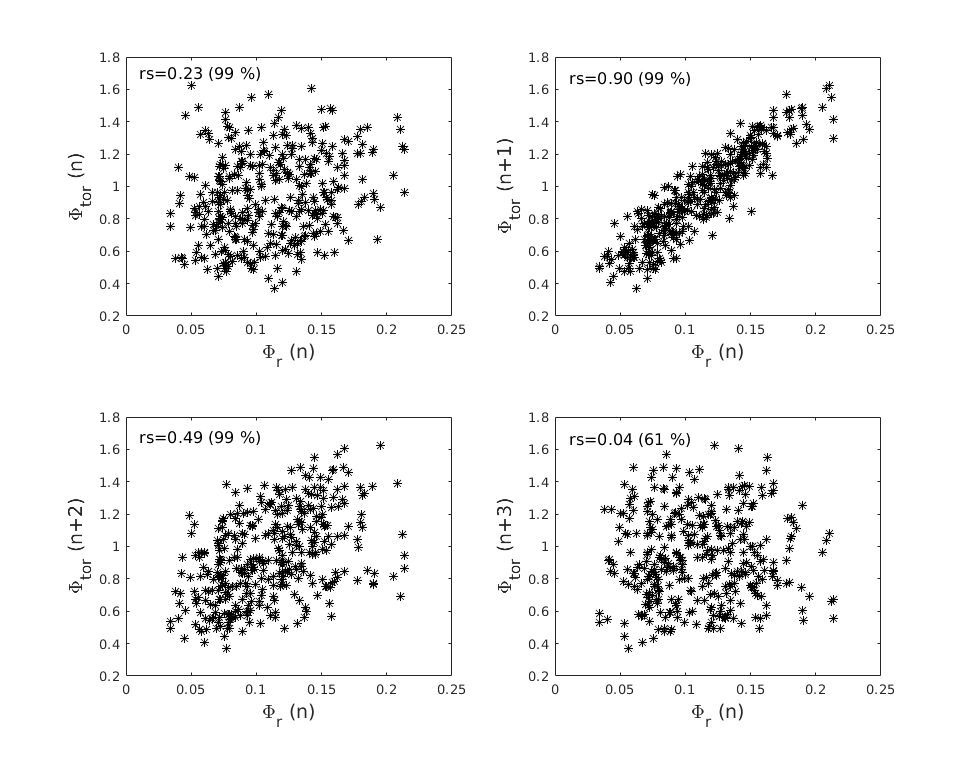}
        \caption{Cross-correlation between the radial flux ($\Phi_r$) of cycle n and the toroidal flux ($\Phi_{tor}$) of cycle n, n+1, n+2, n+3 in the advection-dominated dynamo. The poloidal field is generated only due to the Babcock-Leighton mechanism. Spearman correlation coefficients with significance level are given inside the plots. Here, we consider 60 \% fluctuations in the Babcock-Leighton mechanism.}
\end{figure*}
\begin{figure*}
        \centering
        \includegraphics[width=15 cm]{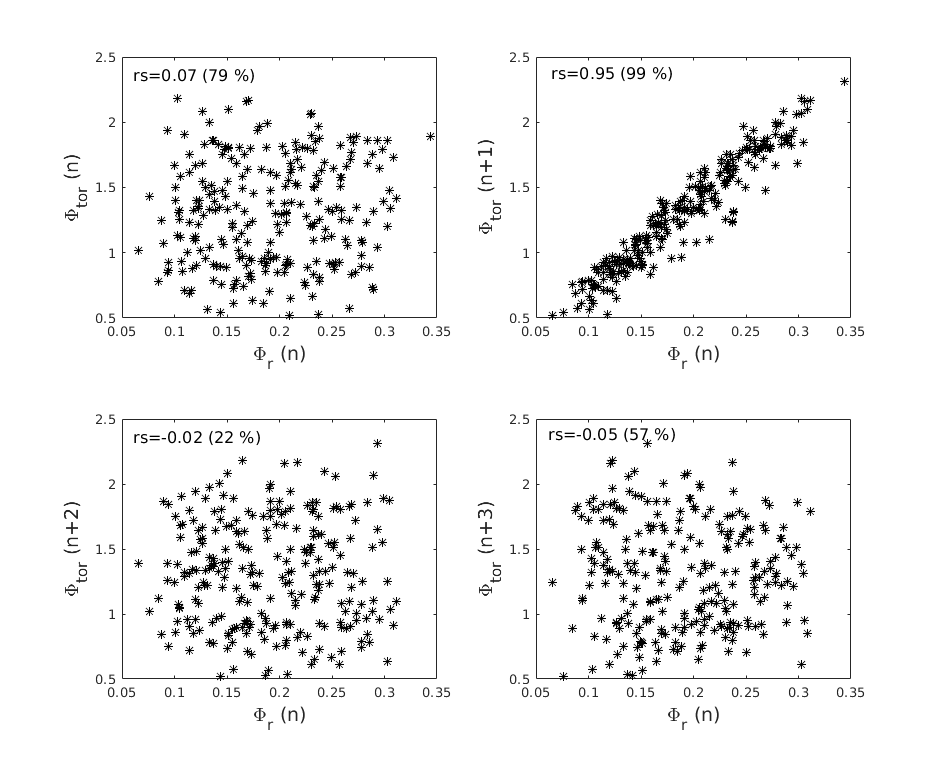}
        \caption{Cross-correlation between the radial flux ($\Phi_r$) of cycle n and the toroidal flux ($\Phi_{tor}$) of cycle n, n+1, n+2, n+3 in the diffusion-dominated dynamo. The poloidal field is generated only due to the Babcock-Leighton mechanism. Spearman correlation coefficients with significance level are given inside the plots. Here, we consider 60 \% fluctuations in the Babcock-Leighton mechanism.}
\end{figure*}
\begin{figure*}
        \centering
        \includegraphics[width=15 cm]{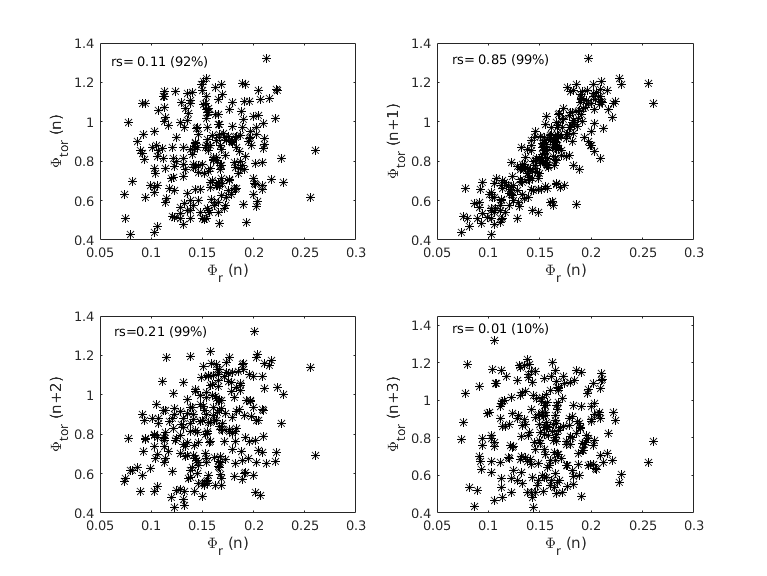}
        \caption{Cross-correlation between the radial flux ($\Phi_r$) of cycle n and the toroidal flux ($\Phi_{tor}$) of cycle n, n+1, n+2, n+3 in the advection-dominated dynamo. The poloidal field is generated due to both the Babcock-Leighton mechanism and mean-field alpha effect. Spearman correlation coefficients with significance level are given inside the plots. Here, we consider 60 \% fluctuations in the Babcock-Leighton mechanism and 50 \% fluctuation in the mean field alpha. The value of the mean-field alpha constant factor $S_\odot$ is 0.14.}
\end{figure*}
\begin{figure*}
        \centering
        \includegraphics[width=15 cm]{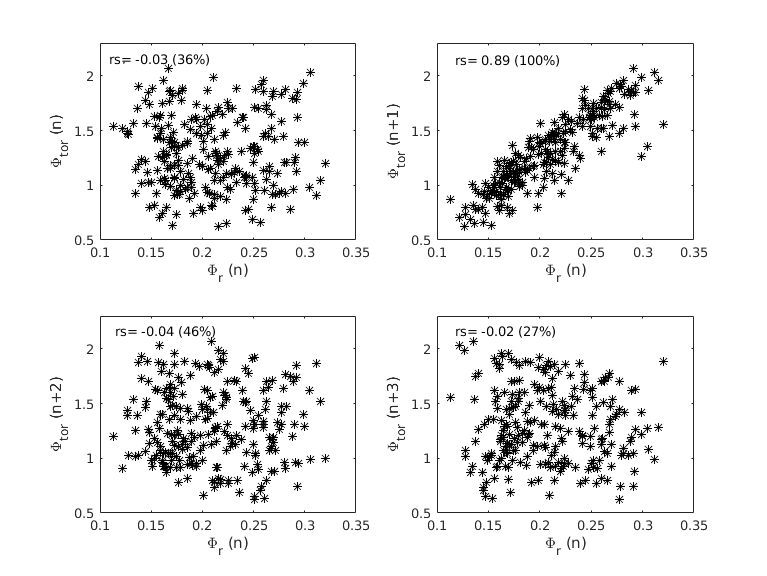}
        \caption{Cross-correlation between the radial flux ($\Phi_r$) of cycle n and the toroidal flux ($\Phi_{tor}$) of cycle n, n+1, n+2, n+3 in the diffusion-dominated dynamo. Poloidal field is generated due to both the Babcock-Leighton mechanism and mean-field alpha effect. Spearman correlation coefficients with significance level are given inside the plots. Here, we consider 60 \% fluctuations in the Babcock-Leighton mechanism and 50 \% fluctuation in the mean field alpha. The value of the mean-field alpha constant factor $S_\odot$ is 0.14.}
\end{figure*}

In the first scenario, we consider the Babcock-Leighton mechanism as the only poloidal field generation mechanism. We first perform the simulation without any fluctuation. We are able to reproduce a solar-like cycle with an 11-year periodicity. We also confirm that the periodicity of the solar cycle decreases with the speed of the meridional flow \citep{dikp99}. Periodicity lies within the range of 7 to 18 years depending on the speed of the meridional flow at the surface, which varies from $25$ m s$^{-1}$ to  $10$ m s$^{-1}$. However, in reality, the Babcock-Leighton mechanism is a random process. The stochastic nature of the Babcock-Leighton mechanism arises from the random buffeting of flux tubes during their rise through the turbulent convection zone, yielding a significant scatter in tilt angles of the active region \citep{long02, mccl16}. Motivated by these facts, we introduce stochastic fluctuation in the poloidal field generation source term by setting up $K_1 = K_{base} + K_{fluc} \sigma(t, \tau_{cor})$ with $K_{base}=100$ in the double-ring algorithm. Here $\sigma$, the uniform random number, lies between -1 and +1. We run all our following simulations with a 60 \% fluctuation in the Babcock-Leighton mechanism, and therefore in our simulation $K_{fluc}= 0.6 K_{base}$ (see Appendix). Our choice of fluctuation level is inspired by observations as well as the eddy velocity distributions present in 3D turbulent convection simulations \citep{mies08, raci11, pass12}. 

\begin{table*}[t]
\caption[]{Correlation coefficients ($r_s$) and percentage significance levels ($p$) for peak surface radial flux $\Phi_{\rm{r}}$ of cycle $n$ versus peak toroidal flux $\Phi_{\rm{tor}}$ of different cycles from 
data for 200 solar cycles. Here, we consider a 60 \% fluctuation in the Babcock-Leighton mechanism but no fluctuation in the mean-field alpha.}
 \begin{center}\begin{tabular}{lrrrrr}
 \hline
                               &              &Dif. Dom.                                       & Adv. Dom.      \\
 \hline
Mean Field Alpha ($S_\odot$)                       & Parameters~~~~~~~                                 & $r_s$ ($p$)& $r_s$~($p$) \\
\hline
                               & $\Phi_{\rm{r}}(n)~\&~\Phi_{\rm{tor}}(n)$~~~~~~& $0.04~(45.0)$ & $-0.29~(99.9)$\\
                               & $\Phi_{\rm{r}}(n)~\&~\Phi_{\rm{tor}}(n+1)$    & $0.91~(99.9)$ & $0.82~(99.9)$\\ [-1ex]
\raisebox{1.0ex}{0.14}         & $\Phi_{\rm{r}}(n)~\&~\Phi_{\rm{tor}}(n+2)$    & $-0.07~(70.0)$ & $0.08~(76.0)$\\
                               & $\Phi_{\rm{r}}(n)~\&~\Phi_{\rm{tor}}(n+3)$    & $-0.04~(42.0)$ & $0.09~(75.9)$\\
\hline
                               & $\Phi_{\rm{r}}(n)~\&~\Phi_{\rm{tor}}(n)$~~~~~~& $0.05~(50.0)$ & $0.41~(99.9)$\\
                               & $\Phi_{\rm{r}}(n)~\&~\Phi_{\rm{tor}}(n+1)$    & $0.85~(99.9)$ & $0.75~(99.9)$\\[-1ex]
\raisebox{1.0ex}{0.20}         & $\Phi_{\rm{r}}(n)~\&~\Phi_{\rm{tor}}(n+2)$    & $-0.08~(68.9)$ & $0.29~(99.9)$\\
                               & $\Phi_{\rm{r}}(n)~\&~\Phi_{\rm{tor}}(n+3)$    & $-0.02~(42.0)$ & $-0.01~(18.9)$\\
\hline
\end{tabular}
  \end{center}
\end{table*}

\begin{table*}[t]
\caption[]{Correlation coefficients ($r_s$) and percentage significance levels ($p$) for peak surface radial flux $\Phi_{\rm{r}}$ of cycle $n$ versus peak toroidal flux $\Phi_{\rm{tor}}$ of different cycles from data for 200 solar cycles. In all cases, we run our simulations with a 60 \% fluctuation in the Babcock-Leighton mechanism. The value of the mean-field alpha constant $S_\odot$ is 0.14 ms$^{-1}$in all cases.}
 \begin{center}\begin{tabular}{lrrrrr}
 \hline
                               &              &Dif. Dom.                                       & Adv. Dom.      \\
 \hline
Percentage fluctuation                        & Parameters~~~~~~~                                 & $r_s$ ($p$)& $r_s$~($p$) \\
in mean field alpha  \\
\hline
                               & $\Phi_{\rm{r}}(n)~\&~\Phi_{\rm{tor}}(n)$~~~~~~& $0.04~(45.0)$ & $-0.29~(99.9)$\\
                               & $\Phi_{\rm{r}}(n)~\&~\Phi_{\rm{tor}}(n+1)$    & $0.91~(99.9)$ & $0.82~(99.9)$\\ [-1ex]
\raisebox{1.0ex}{0}         & $\Phi_{\rm{r}}(n)~\&~\Phi_{\rm{tor}}(n+2)$    & $-0.07~(70.0)$ & $0.08~(76.0)$\\
                               & $\Phi_{\rm{r}}(n)~\&~\Phi_{\rm{tor}}(n+3)$    & $-0.04~(42.0)$ & $0.09~(76.0)$\\
\hline
                               & $\Phi_{\rm{r}}(n)~\&~\Phi_{\rm{tor}}(n)$~~~~~~& $0.11~(79.0)$ & $-0.08~(86.1)$\\
                               & $\Phi_{\rm{r}}(n)~\&~\Phi_{\rm{tor}}(n+1)$    & $0.91~(99.9)$ & $0.76~(99.9)$\\ [-1ex]
\raisebox{1.0ex}{10}         & $\Phi_{\rm{r}}(n)~\&~\Phi_{\rm{tor}}(n+2)$    & $0.04~(33.0)$ & $0.29~(97.9)$\\
                               & $\Phi_{\rm{r}}(n)~\&~\Phi_{\rm{tor}}(n+3)$    & $0.15~(75.0)$ & $0.10~(80.0)$\\
\hline
                               & $\Phi_{\rm{r}}(n)~\&~\Phi_{\rm{tor}}(n)$~~~~~~& $0.11~(61.0)$ & $-0.11~(91.9)$\\
                               & $\Phi_{\rm{r}}(n)~\&~\Phi_{\rm{tor}}(n+1)$    & $0.89~(99.9)$ & $0.78~(99.9)$\\ [-1ex]
\raisebox{1.0ex}{20}         & $\Phi_{\rm{r}}(n)~\&~\Phi_{\rm{tor}}(n+2)$    & $0.12~(67.0)$ & $0.20~(98.0)$\\
                               & $\Phi_{\rm{r}}(n)~\&~\Phi_{\rm{tor}}(n+3)$    & $0.22~(94.0)$ & $0.04~(43.0)$\\

\hline
                               & $\Phi_{\rm{r}}(n)~\&~\Phi_{\rm{tor}}(n)$~~~~~~& $0.02~(23.08)$ & $0.16~(97.9)$\\
                               & $\Phi_{\rm{r}}(n)~\&~\Phi_{\rm{tor}}(n+1)$    & $0.89~(99.9)$ & $0.81~(99.9)$\\[-1ex]
\raisebox{1.0ex}{30}         & $\Phi_{\rm{r}}(n)~\&~\Phi_{\rm{tor}}(n+2)$    & $-0.08~(79.2)$ & $0.37~(99.9)$\\
                               & $\Phi_{\rm{r}}(n)~\&~\Phi_{\rm{tor}}(n+3)$    & $-0.02~(21.3)$ & $0.16~(96.4)$\\
\hline
                               & $\Phi_{\rm{r}}(n)~\&~\Phi_{\rm{tor}}(n)$~~~~~~& $0.13~(73.0)$ & $0.11~(91.9)$\\
                               & $\Phi_{\rm{r}}(n)~\&~\Phi_{\rm{tor}}(n+1)$    & $0.88~(99.9)$ & $0.74~(99.9)$\\ [-1ex]
\raisebox{1.0ex}{40}         & $\Phi_{\rm{r}}(n)~\&~\Phi_{\rm{tor}}(n+2)$    & $-0.03~(19.0)$ & $0.33~(97.9)$\\
                               & $\Phi_{\rm{r}}(n)~\&~\Phi_{\rm{tor}}(n+3)$    & $-0.01~(8.0)$ & $0.21~(98.0)$\\                               
\hline
                               & $\Phi_{\rm{r}}(n)~\&~\Phi_{\rm{tor}}(n)$~~~~~~& $-0.03~(36.0)$ & $0.11~(92.0)$\\
                               & $\Phi_{\rm{r}}(n)~\&~\Phi_{\rm{tor}}(n+1)$    & $0.89~(99.9)$ & $0.85~(99.9)$\\[-1ex]
\raisebox{1.0ex}{50}         & $\Phi_{\rm{r}}(n)~\&~\Phi_{\rm{tor}}(n+2)$    & $-0.04~(46.0)$ & $0.21~(99.0)$\\
                               & $\Phi_{\rm{r}}(n)~\&~\Phi_{\rm{tor}}(n+3)$    & $-0.02~(27.0)$ & $0.01~(10.0)$\\
\hline
                               & $\Phi_{\rm{r}}(n)~\&~\Phi_{\rm{tor}}(n)$~~~~~~& $0.22~(93.0)$ & $0.05~(38.9)$\\
                               & $\Phi_{\rm{r}}(n)~\&~\Phi_{\rm{tor}}(n+1)$    & $0.84~(99.9)$ & $0.61~(99.9)$\\ [-1ex]
\raisebox{1.0ex}{60}         & $\Phi_{\rm{r}}(n)~\&~\Phi_{\rm{tor}}(n+2)$    & $-0.02~(16.0)$ & $0.18~(93.0)$\\
                               & $\Phi_{\rm{r}}(n)~\&~\Phi_{\rm{tor}}(n+3)$    & $0.15~(77.0)$ & $0.09~(60.0)$\\                               
\hline
\end{tabular}
  \end{center}
\end{table*}
The middle panel of Fig.~1 shows the butterfly diagram at the base of the convection zone from a simulation in the diffusion-dominated regime and the bottom panel shows the variation of $B_\phi^2$ (a proxy of sunspot number) with time at the base of the convection zone. We note that the sensitivity of the peak amplitude of the $B_\phi^2$ is due to the choice of fluctuation level. We calculate the polar radial flux $\Phi_r$ and toroidal flux $\Phi_{tor}$ using the prescription suggested by \cite{kara12} and \cite{yeat08}. The toroidal flux $\Phi_{tor}$ is calculated by integrating $B_\phi(r, \theta)$ within a layer of $r= 0.677 R_\odot - 0.726 R_\odot$ and within the latitude $10^{\circ}- 45^{\circ}$; while the radial flux $\Phi_r$ is calculated by integrating $B_r (R_\odot, \theta)$ at the solar surface within the latitude $70^{\circ}-89^{\circ}$. We note that there is a $90^{\circ}$ phase difference between the radial flux and the toroidal flux. Radial flux is maximum at the minima of the solar cycle. We find the peak value of $\Phi_r$ and $\Phi_{tor}$ for each cycle and study the cross-correlation between the surface radial flux $\Phi_r$ of cycle n and the toroidal flux of cycle n, n+1, n+2, and n+3. We perform the same study for both advection-dominated and diffusion-dominated dynamo simulations. Cycle-to-cycle correlation gives us the extent of correlation between the radial and toroidal flux. As per suggestions by \cite{yeat08} and \cite{kara12}, the extent of the correlation is an indicator of the memory of the solar cycle. We run our stochastically forced dynamo model for a total of 250 solar cycles to generate the correlation statistics. 

Figure 2 shows that in the advection-dominated dynamo, surface radial flux $\Phi_r$ correlates with the toroidal flux $\Phi_{tor}$ of cycles (n+1) and (n+2) with Spearman correlation coefficients 0.90 and 0.49, respectively. On the other hand, Figure 3 indicates that in the diffusion-dominated dynamo, surface radial flux $\Phi_r$ only correlates with the toroidal flux $\Phi_{tor}$ of the subsequent cycle (correlation coefficient 0.95). \cite{yeat08} studied the memory of the solar cycle and found that in the diffusion-dominated dynamo surface radial flux $\Phi_r$ only correlates with the subsequent cycle toroidal flux $\Phi_{tor}$. While, in the advection-dominated dynamo, surface radial flux $\Phi_r$ correlates with the toroidal flux $\Phi_{tor}$ of subsequent few cycles (n+1), (n+2), and (n+3). \cite{yeat08} and \cite{kara12} found higher correlation coefficients in the advection-dominated regimes than those that we present here. This is probably due to our choice of modelling magnetic buoyancy by the double-ring algorithm. In summary, our results agree with the results of \cite{yeat08} when we consider the Babcock-Leighton mechanism as the only poloidal-field-generation process.

In the second scenario, we consider both the Babcock-Leighton mechanism and mean-field alpha effect as poloidal-field-generation mechanisms. However, the mean-field alpha effect is also a random process, not a deterministic one.  As the mean-field alpha effect arises due to helical turbulence inside the turbulent convection zone, the mean-field alpha effect is also a stochastic process. Motivated by these factors, we introduce randomness into the mean-field alpha by setting $S_\odot= S_{base} + S_{fluc} \sigma(t, \tau_{corr})$. Here, $\sigma$ is a uniform random number lying between $-1$ and $+1$. We set the correlation time $\tau_{corr}$ in such a way that at least ten fluctuations are there within a single solar cycle. We run our simulations with a 60 \% fluctuation in the Babcock-Leighton mechanism and a different level of fluctuation in the mean-field alpha effect. Figures 4 and 5 show that in the case of both advection- and diffusion-dominated dynamos, surface radial flux $\Phi_r$ only correlates with the toroidal flux of the subsequent cycle with Spearman correlation coefficients of 0.85 and 0.89, respectively. Please note that we use the results of the model with 60 \% fluctuation in the Babcock-Leighton mechanism and 50 \% fluctuation in the mean-field alpha effect to generate Figures 4 and 5. Tables 1 and 2 summarise results from parameter space studies with variations in the amplitude and fluctuation levels of the mean-field alpha effect. In our model, the presence of the mean-field alpha effect reduces the memory of the solar cycle to only one cycle if there is no fluctuation in the mean-field alpha. However, when fluctuations are introduced in the mean-field alpha effect, very weak but significant correlations emerge between the radial flux at the n-th cycle minima and the toroidal flux at the (n+2)-th cycle maxima for the advection-dominated dynamo (see Table 2). Table 1 shows that the correlation strength between the radial and toroidal flux of various cycles are also dependent on the assumed amplitude of the mean-field alpha effect (which is restricted to a value of $0.20$ ms$^{-1}$ for reasons of stability). The important conclusion from this study is that we get only one cycle memory even in the case of the advection-dominated convection zone for steady mean-field alpha (no fluctuations) and that variation in the amplitude or fluctuation levels of mean-field alpha effect has a small but measurable impact on the memory of the advection-dominated dynamo mechanism.
 
\section{Conclusions}
 
In summary, we demonstrate that inclusion of the mean-field alpha effect has a measurable impact on the memory of the solar cycle. We find that solar-cycle memory is only limited to one cycle when a basal steady mean-field alpha effect is included in a flux-transport-type dynamo model and acts in conjunction with the Babcock-Leighton mechanism. This result supports earlier suggestions regarding the precursor value of the poloidal field at the end of a cycle for the amplitude of the subsequent cycle alone \citep{scha78, sola02, yeat08, kara12, muno13, sanc14}. Our detailed analysis shows that the presence of a mean-field alpha effect in the dynamo model adds more complexity to the interpretation of differences in the memory of the solar dynamo in the context of advection- versus diffusion-dominated solar convection zones. Specifically, we find that variations in the amplitude or fluctuation levels of the mean-field alpha can have a weak but measurable impact on the emergence of multi-cycle memory in the advection-dominated solar dynamo.

We do not consider turbulent pumping in our simulations. \cite{kara12} show that turbulent pumping can impact the memory of the solar cycle. These latter authors found that in the case of both advection- and diffusion-dominated regimes, surface radial flux $\Phi_r$ only correlates with the subsequent cycle toroidal flux if turbulent pumping is added to their model. However, \cite{kara12} did not include the mean-field alpha effect in their dynamo simulations. We also find that surface radial flux $\Phi_r$ correlates only with the toroidal flux $\Phi_{tor}$ of the subsequent cycle when we include turbulent pumping in our simulations with mean-field alpha effect which extends the validity of the earlier results. It has been argued that the relative efficiency between different flux transport mechanisms governs the memory of the solar cycle. In the model that considers the Babcock-Leighton mechanism as the sole poloidal field generation mechanism, the timescale for transport of the poloidal flux from the surface to the base of the convection zone governs the memory of the solar cycle. Even for a modest radial turbulent pumping speed of $2$ ms$^{-1}$, the timescale for transport of the poloidal flux from surface to the base of the convection zone is only 3.4 years. The introduction of turbulent pumping in the flux transport dynamo model makes the dynamo model completely dominated by turbulent pumping, eventually impacting the memory of the solar cycle. In the situation where turbulent pumping dominates the vertical flux transport mechanism, the relative efficiency of other flux transport mechanisms, namely meridional circulation and turbulent diffusion, is less significant.

In our model, the poloidal field generation takes place due to the combined effect of both the Babcock-Leighton mechanism and the mean-field alpha effect. As poloidal field generation takes place throughout the convection zone due to the mean-field alpha effect, the poloidal flux becomes immediately available for being inducted to the toroidal component by the differential rotation.  In the absence of any additional complexities, this would normally result in a reduction of the cycle memory to only one cycle -- as seen in our case with no fluctuation in mean-field alpha. However, introduction of fluctuations in the mean-field alpha effect and variations in the relative strengths of the competing mean-field and Babcock-Leighton poloidal sources may result in a redistribution of the memory across cycles in the advection-dominated regime; we speculate that this is at the origin of the emergence of weak multi-cycle correlations and their variations with increasing levels of fluctuation in the mean-field alpha effect.  

Taken together, we argue that additional consideration of a mean-field alpha effect within the framework of a flux transport dynamo model -- driven by a Babcock-Leighton mechanism -- introduces subtleties in the dynamics that may be weak, but may still be important to capture from the perspective of predictive dynamo models. Predictive mode coupled surface flux transport and dynamo simulations support this argument in the context of the successful reconstruction of past solar activity cycles \citep{bhow18}.
Completely independent considerations of cycle recovery following a grand minimum in activity also support the importance of a mean-field alpha effect in the bulk of the solar convection zone -- even if it is weak in strength \citep{shaz14,pass14}. We therefore conclude that it may be necessary to include the physics of the mean-field alpha effect in predictive dynamo models of the solar cycle, even if the Babcock-Leighton mechanism are the dominant source for poloidal field generation. 

\section{Appendix: Modelling the Babcock-Leighton mechanism}
Poloidal field generation at the surface takes place due to the decay and dispersal of the bipolar sunspot regions as well as ephemeral regions. We model the poloidal field generation mechanism at the surface due to ephemeral regions following an equation similar to Eq. (8):
\begin{eqnarray}
    S_{EPR}= S_{1} \frac{ \cos \theta }{4} \left[1+\textrm{erf}
    \left( \frac{r-r_1}{d_1}\right)\right]  
    \left[1-\textrm{erf}\left( \frac{r-r_2}{d_2}\right)\right] \nonumber \\ 
    \times \frac{1}{1+\left(\frac{B_\phi}{B_{up}}\right)^2} ~~~~
   ,\end{eqnarray}
 where we take $r_1=0.95 R_0$, $r_2=R_0$, $d_1=d_2=0.15 R_0$, and $B_{up}= 10^4~G$. Here, $S_1$ controls the amplitude of the poloidal field generation mechanism due to ephemeral regions. We take $S_1=0.13$ in our model.
 
In our model, we follow the prescription of the double-ring algorithm proposed by \cite{durn97} to model the active region. In this algorithm,  we define the $\phi$ component of potential vector A associated with the active region as:
  \begin{equation}\label{Eq_AR}
    A_{ar}(r,\theta)= K_1 A(\Phi)F(r)G(\theta),
\end{equation}
where constant $K_1$ ensures the super-critical dynamo solution and $A(\Phi)$ defines the strength of the ring doublet. $F(r)$ is defined as:
\begin{equation}
    F(r)= \left\{\begin{array}{cc}
            0 & r<R_\odot-R_{ar}\\
            \frac{1}{r}\sin^2\left[\frac{\pi}{2 R_{ar}}(r - (R_\odot-R_{ar}))\right] & r\geq R_\odot-R_{ar}
          \end{array}\right.,
\end{equation}
where $R_\odot$ is the solar radius and the penetration depth of the active region is $R_{ar}=0.85R_\odot$. $G(\theta)$  in the integral form is defined as:
\begin{equation}
    G(\theta) = \frac{1}{\sin{\theta}}\int_0^{\theta}[B_{-}(\theta')+B_{+}(\theta')]\sin(\theta')d\theta',
\end{equation}
where $B_{+}$ ($B_{-}$) represents the strength of positive (negative) ring:
 \begin{eqnarray}\label{Eq_AR_Dp}
    B_{\pm}(\theta)= \left\{\begin{array}{cc}
                     0 & \theta<\theta_{ar}\mp\frac{\chi}{2}-\frac{\Lambda}{2}\\
                     \pm\frac{1}{\sin(\theta)}\left[1+\cos\left(\frac{2\pi}{\Lambda}(\theta-\theta_{ar}\pm\frac{\chi}{2})\right)\right] \\
                     \theta_{ar}\mp\frac{\chi}{2}-\frac{\Lambda}{2} \leq \theta < \theta_{ar}\mp\frac{\chi}{2}+\frac{\Lambda}{2}\\
                     0 & \theta \geq \theta_{ar}\mp\frac{\chi}{2}+\frac{\Lambda}{2},
               \end{array}\right.
\end{eqnarray}
where $\theta_{ar}$ is the colatitude of the double ring emergence and the diameter of each polarity of the double ring is $\Lambda$. We take the latitudinal distance between the centres of the double ring as $\chi = \arcsin[\sin(\beta)\sin(\Delta_{ar})]$, where $\Delta_{ar}$ is the angular distance between polarity centres and $\beta$ is the active region tilt angle.  We take $\Lambda$ and $\Delta_{ar}$ as $6^o$ for our model.\\

{\bf Regenerating the poloidal field}:

To recreate the poloidal field at the solar surface, we first randomly choose  a latitude from both the hemispheres where the toroidal field exceeds the buoyancy threshold at the bottom of the convection zone. We then use a non-uniform probability distribution function to ensure that randomly chosen latitude always remains within the observed active region belt. Next, we calculate the tilt of the corresponding active region following the expression prescribed in \cite{fan94}:
\begin{equation}\label{Eq_Prob}
   \beta \propto \Phi_0^{1/4}B_0^{-5/4}\sin(\lambda),
\end{equation} 
where, $\Phi_0$ is the toroidal ring associated flux, $B_0$ is the local field strength, and $\lambda$ is the chosen latitude for the ring emergence. The constant that appears in equation 14 is fixed in a way such that the tilt angle lies between  3$^{\circ}$ and  12$^{\circ}$. Next, we remove a part of the magnetic field with the same angular size of the emerging active region from this toroidal ring. We reduced the magnetic field strength of the toroidal ring from which the active region erupts. We set the toroidal field strength in  such a way that the energy of the full toroidal ring with the new magnetic field is equal to the energy of the partial toroidal ring with the old magnetic field (after removing a chunk of the magnetic field). Finally, we place the ring doublets with these calculated properties at the near-surface layer at the chosen erupted latitude. Figure 1 (right side-top panel) shows the poloidal field line contours associated with the double ring in both hemispheres for one particular time-step.

\begin{acknowledgements}
We thank Antoine Strugarek and Eric Buchlin for reading this manuscript and providing useful suggestions. We also thank the anonymous referee for the constructive suggestions which have improved the work further. We thank the Université Paris-Saclay (IRS SPACEOBS grant), ERC Synergy grant 810218 (Whole Sun project), INSU/PNST and CNES Solar Orbiter for supporting this research. The Center of Excellence in Space Sciences India (CESSI) is supported by the Ministry of Human Resource Development, Government of India under the Frontier Areas of Science and Technology (FAST) scheme. 
\end{acknowledgements}








\end{document}